\documentclass{article}

\usepackage{PRIMEarxiv}

\usepackage[utf8]{inputenc} 
\usepackage[T1]{fontenc}    
\usepackage{hyperref}       
\usepackage{url}            
\usepackage{booktabs}       
\usepackage{amsfonts}       
\usepackage{nicefrac}       
\usepackage{microtype}      
\usepackage{fancyhdr}       
\usepackage{graphicx}       

\usepackage{amsmath,amssymb}
\usepackage{comment}
\usepackage{textcomp,marvosym}
\usepackage{nameref}
\usepackage{array}
\usepackage{relsize}
\usepackage{tabularray}

\title{Investor-patent networks as mutualistic networks
}

\author{
  Théophile Carniel \\
  Université Paris Cité, CNRS, LIED UMR 8236, F-75006 Paris, France \\
  Agoranov, Paris, France \\
  \\
   \And
  Léo Cazenille \\
  Université Paris Cité, CNRS, LIED UMR 8236, F-75006 Paris, France\\
  \\
   \And
  Jean-Michel Dalle \\
  Agoranov, Paris, France \\
  Sorbonne Université, Paris, France \\
  \\
   \And
  José Halloy \\
  Université Paris Cité, CNRS, LIED UMR 8236, F-75006 Paris, France \\
  \texttt{jose.halloy@u-paris.fr} \\
}

\begin{document}
\maketitle

\begin{abstract}
Venture capital investments in startups have come to represent an important driver of technological innovation, in parallel to corporate- and government-directed efforts. Part of the future of artificial intelligence, medicine and quantum computing now depends upon a large number of venture investment decisions whose robustness against increasingly frequent crises has therefore become crucial. To shed light on this issue, and by combining large-scale financial, startup and patent datasets, we analyze the interactions between venture capitalists and technologies as an explicit bipartite patent-investor network. Our results reveal that this network is topologically mutualistic because of the prevalence of links between generalist investors, whose portfolios are technologically diversified, and general-purpose technologies, characterized by a broad spectrum of use. As a consequence, the robustness of venture-funded technological innovation against different types of crises is affected by the high nestedness and low modularity, with high connectance, associated with mutualistic networks.
\end{abstract}


\section{Introduction}
Due to the increased role of startups in various technological fields, from biotechnology to artificial intelligence or quantum computing~\cite{audretsch2020innovative}, venture capital has \textit{de facto} become an important driver of technological innovation~\cite{kortum2000assessing, ferrary2017social, ferrary2009role, lerner2020venture}. Indeed, because of their inability to self-finance during the early years of their operations, startups need to rely on  the investments that they themselves receive from specialized investors called venture capitalists (VCs). In this context, the relationships between VC funding and innovation have gradually become a topic of interest, mostly approached through patent data~\cite{nanda2013investment, gonzalez2020exchanges, howell2020resilient, dalla2023innovating}. This evolution has occurred in a world where crises have become more and more frequent~\cite{harris2020law}, increasing the need to analyze the resilience of socio-economic systems~\cite{frank2019toward}. These studies have notably preliminarily shown that VC funding could be negatively impacted by local and global crises, be they financial~\cite{block2010venture}, health~\cite{bellavitis2021covid} or geopolitical~\cite{kramer2023entrepreneurial}.
However, it is quite surprising that, although the VC network has been an active topic of study for the past 15 years~\cite{hochberg2007whom, li2024evolution}, and although~\cite{ferrary2009role} had pioneered a complex network approach focused on robustness, the direct interactions between VCs and the innovations they fund, based explicitly on the patents filed~\cite{acemoglu2016innovation} by the startups funded, have not been explicitly studied, both in their own right and in relation to the robustness of the network they constitute.
This VC-patent interaction network is bipartite, with nodes of a first class (VCs) interacting with nodes of a second class (patents) through investments in startups that file the patents.

In this study, considering the line of analysis of financial markets suggested by~\cite{levin2015new,levin2021introduction} and echoing also the approaches that have started to directly address economic complexity~\cite{hidalgo2021economic}, we combine large-scale financial datasets on the rounds of VC funding received by startups with patent data to explicitly analyze this bipartite investor-patent network and its emergent structure.
We identify clusters of investors and clusters of patents and observe that their bipartite network is topologically mutualistic, \textit{i.e.} that the structure of the network shares metric characteristics with mutualistic networks in ecology. This is due to the prevalence of investors whose financial incentives make them diversify their portfolios with respect to technological innovations in order to reduce the risks taken~\cite{buchner2017diversification} and to the existence of a large number of general-purpose technologies, \textit{i.e.} technologies whose use spreads widely across economic sectors~\cite{jovanovic2005general}.

With respect to the robustness of this network, we analyze its nestedness~\cite{bascompte2003nested} and modularity~\cite{beckett2016improved} metrics, as they have been developed by the ecological and physical literature. Nestedness measures the existence of a matryochka-like structure of interaction, where specialist nodes interact with nodes that the generalist nodes also interact with. Modularity estimates the propensity of nodes in a module (a set of nodes allocated to the same group) to interact with nodes in the same module. Both metrics have been linked to the system response to perturbations. We find the investor-patent network to be strongly nested and weakly modular, which is consistent with its topologically mutualistic nature. As a consequence, this network is characterized by distinct responses to different system perturbations~\cite{thebault2010stability}. Crises that affect investors randomly or specialized investors tend to have relatively little impact, due to the redundant nested structure of the network, whereas events that target generalist investors tend to present a higher risk for related technological innovations~\cite{burgos2007nestedness}.

\section*{Objectives}
We study the interactions between investors and the technologies developed by startup companies in their portfolios. To do so, we combine financial and patent data in a network analysis framework. We first present the methodology used to build this bipartite network. We then analyze its topology and, using metrics developed by the literature in ecology and physics, notably nestedness and modularity, we discuss the implications of this topology for its robustness against crises.

\section{Materials and methods}

\begin{figure*}[ht]
    \centering
    \includegraphics[width=0.90\textwidth]{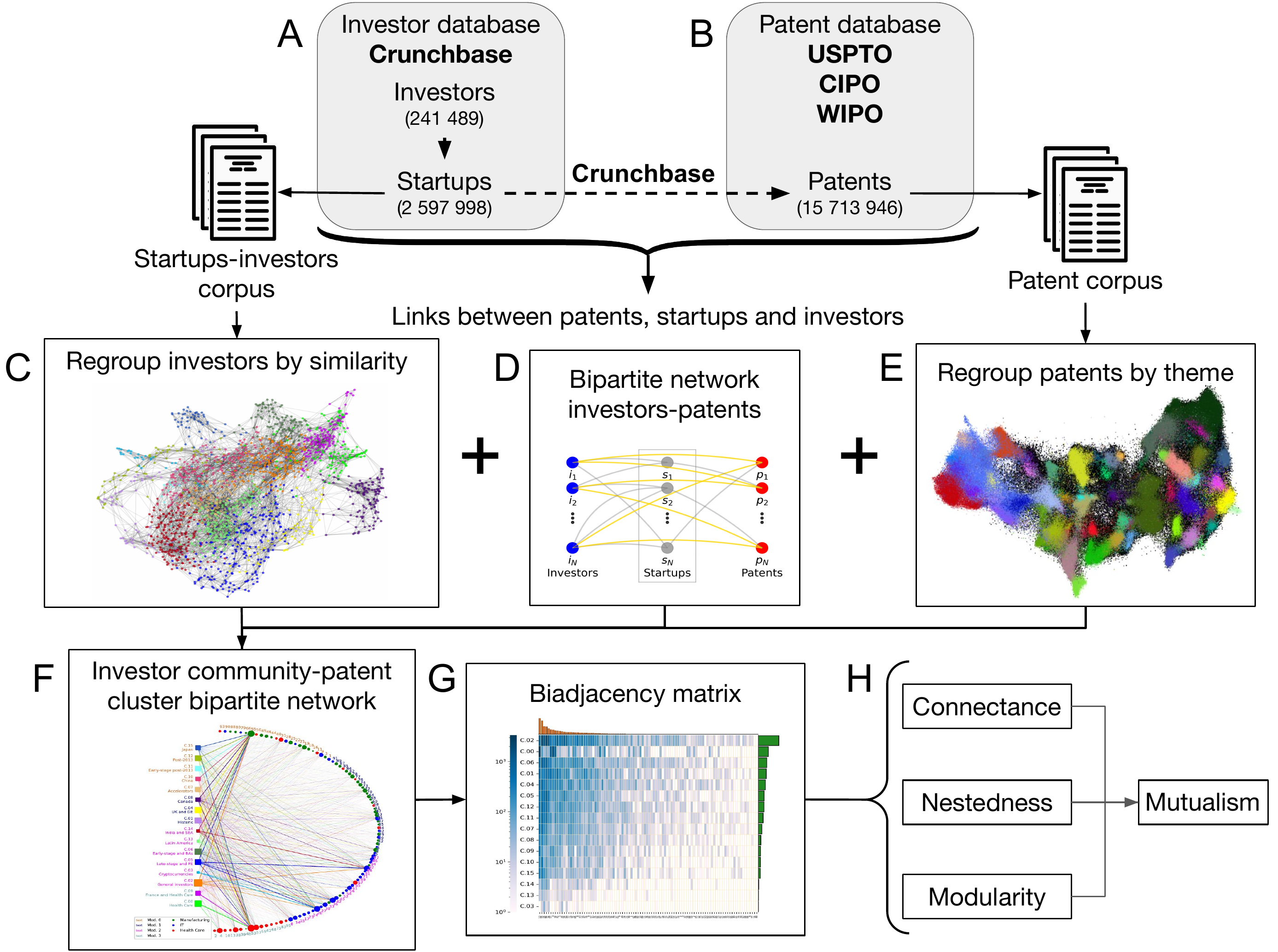}
    \caption{Workflow presenting the approach followed in this article.
    \textbf{(A)} Investor and company information is extracted from Crunchbase. \textbf{(B)} Patent data from USPTO, CIPO and WIPO is extracted and matched with the Crunchbase company information. 
    \textbf{(C)} $16$ investors communities are retrieved using a similarity metric between pairs of investors. 
    \textbf{(D)} The bipartite network between investors and the patents of the companies they invested in is built. 
    \textbf{(E)} NLP-based topic modeling of patents is performed on their abstracts and $98$ patent clusters are retrieved. 
    \textbf{(F)} The investor community-patent cluster graph is built based on the investor-patents bipartite network by combining the results from steps \textbf{(C)}, \textbf{(D)} and \textbf{(E)}, \textit{i.e.} by aggregating investors into their investor communities and patents into their patent clusters on the investor-patents bipartite network. 
    \textbf{(G)} The biadjacency matrix of the investor community-patent cluster graph is extracted to quantitatively visualize the interaction patterns and compute network structure metrics. 
    \textbf{(H)} Network structure metrics (connectance, nestedness, modularity) are computed using the biadjacency matrix to study the topology of the network and the properties deriving from it.}
    \label{fig:workflow}
\end{figure*}

Figure~\ref{fig:workflow} presents the approach followed in this article. Using financial and patent databases, we retrieve information concerning startups, the patents they own and the investors that financed them. Investor communities are then identified using a pairwise investor similarity methodology from the literature and patent clusters are identified using Topic Modeling methods. These aggregate-level descriptions are then used to build a bipartite graph linking investor communities to patent clusters by using the startups as the bridge between the two (startups can be linked both to the patents they own and to the investors that financed them). We then use the biadjacency matrix of the network to quantitatively investigate its structure through network-level structural metrics frequently used in the ecological literature.

\subsection{Datasets}
\label{subsec:datasets}
The startup dataset used for this study was extracted through the Crunchbase\footnote{\url{https://www.crunchbase.com}} API on February 14, 2023. It contains information on $2\ 597\ 998$ startups (name, headquarters location, creation date, sectoral tags), $396\ 506$ funding events (funded startup, date of the funding round, investors involved, funding amount, investment stage), $241\ 489$ investors (name, creation date, investor type, headquarter location) and $1\ 631\ 627$ individuals (name, professional experiences, academic education, company board memberships and advisory roles). We removed the \textit{Software} and \textit{Other} sectors from the 47 original sectoral tags as they were found to be redundant, highly non-specific and over-represented (representing a combined total of roughly $13\%$ of all tags in the dataset, first and fourth tags in terms of number of occurrences). We filtered out all companies founded before January 1\textsuperscript{st}, 1998 to remove all companies that were not startups and all companies for which geographical information was not available. Funding rounds that were not VC funding (such as debt financing or grants) were also filtered out as they are carried by other actors than VCs. Since we focus on the interactions between investors and technological innovation, using companies as linking agents between both, we filtered out all companies that did not raise funds. After applying these filters, $234\ 358$ companies remained in our dataset.

The patent dataset also supplied by Crunchbase and IPqwery\footnote{\url{https://ipqwery.com}}) contains a total of $15\ 713\ 946$ patents from WIPO\footnote{World Intellectual Property Organization}, USPTO\footnote{United States Patent and Trademark Office} and CIPO\footnote{Canadian Intellectual Property Office} with their title, abstract, filing date, owner identification, and International Patent Classification (IPC) codes. It provides a matching with the startup dataset that links patent owner IDs to Crunchbase company IDs, allowing us to determine the patent portfolios of startups. We filtered out all patents filed before January 1\textsuperscript{st}, 2000 and all patents that were not owned by companies from our filtered startup dataset, resulting in a final dataset of $835\ 763$ patents.

\subsection{Networks}
\label{subsec:investor-patent_network}

To study the structure of interactions between investors and technological innovations, we consider a network where investors interact with technologies based on their investments in startups and on the patents owned by startups : simply, an investor and a patent are linked if the investor has financed the startup owning the patent. At this level of granularity, the interactions between patents and investors are however too sparse to study fundamental behaviors. Since individual patents are known to belong to classes or clusters associated with different fields of technological innovation~\cite{acemoglu2016innovation}, and since investors belong to different types~\cite{carniel2023novel}, we cluster investors and patents in order to aggregate them into coarser-grained communities.

\subsubsection{Investor communities}
\label{subsubsec:investor_communities}
We detect investor communities following the methodology described in~\cite{carniel2023novel}. We select investors with 60 or more investments bringing the number of investors down to $2017$ and, for each of them, build the $5$ characteristic distributions as described in~\cite{carniel2023novel} : temporal, amount, geographic, series and sectoral investment distributions. We compute a similarity metric between all pairs of investors to build a weighted similarity network where all investors are linked and the edge weights correspond to the pairwise similarities. As all nodes are linked to each other in a pairwise similarity network, the network is then pruned to reduce link density (removing weak links to transform our highly mixed community structure into a simpler, lowly mixed community structure~\cite{kim2022link}) to run a community detection algorithm~\cite{blondel2008fast}. This yields $c = 16$ investor communities as shown in Fig.~\ref{fig:networks}\textbf{A}. The community results are in line with those presented in~\cite{carniel2023novel}, with novel communities emerging as the dataset used was extracted more recently.~\cite{carniel2023novel} have shown that the investor communities provided by this methodology are stable with regard to perturbations to the characteristic distributions of individual investors, suggesting that the underlying investor clustering is robust.

\subsubsection{Patent clusters}
\label{subsubsec:patent_groups}
We apply topic modeling to our patent dataset following the methodology described in~\cite{grootendorst2022bertopic,carniel2022using}. In order to thematically cluster similar patents together based on their textual abstracts, we create vector representations (embeddings) of individual patents using the PatentSBERT~\cite{bekamiri2021patentsberta} model specialized in patent modeling. Each patent is thus represented by a 768-dimensional embedding. We then create a low-dimensional representation of all embeddings using parametric UMAP~\cite{sainburg2021parametric}. Using these UMAP vectors in conjunction with HDBSCAN~\cite{mcinnes2017hdbscan}, a density-based spatial clustering algorithm, we perform the clustering of individual patents. The algorithm works in two phases : first, a clustering is performed by identifying regions of high density and grouping the points in these regions together and a hierarchical approach is then taken to return a flat clustering able to take into account the variable cluster densities. As spatial density-based algorithms perform better when the dimensionality of the data is not too high, the dimensionality reduction step is performed to improve the performance of the HDBSCAN clustering. This process results in $p = 98$ patent clusters. To characterize the patent clusters, we perform a keyword extraction procedure for each cluster using c-TF-IDF, extracting the n-grams that are representative of each patent cluster.

\subsubsection{Investor-patent network}

We allocate each startup to the patent cluster most represented in its patent portfolio. Investors that did not invest in any startup that holds a patent and startups that do not own patents are removed, resulting in $1937$ investors and $12\ 007$ startups after filtering. We then build the weighted investor-patent network based on the funding events between investors and startups : the weight of the edge connecting an investor community and a patent cluster corresponds to the number of times members of the investor community have invested in startups allocated to the patent cluster. The resulting weighted investor-patent network is bipartite since it is composed of two different classes of nodes, with nodes of one class (in our case, investor communities) being only linked to nodes of the other class (here, patent clusters). Other examples of bipartite networks in socio-economic research include the country-product network~\cite{hidalgo2009building} or the country-food production network~\cite{tu2016data}. Specific metrics have been developed to characterize bipartite networks~\cite{valdovinos2019mutualistic} that are presented in the following section.

\subsection{Network metrics}
\subsubsection{Nestedness}
\label{subsec:nestedness}
Nestedness~\cite{bascompte2003nested} is a structural property of bipartite networks that characterizes to what extent specialist species interact with subsets of the species generalist species interact with, meaning that in nested networks, specialist-specialist interactions are infrequent. Mutualistic networks, (\textit{i.e.} networks where both species involved in an interaction have a net benefit such as plant-pollinator or seed dispersal networks have been shown to be significantly nested~\cite{bascompte2003nested}. Possible explanations such as system tolerance to species extinctions have been suggested as a reason for the prevalence of nestedness in mutualistic systems~\cite{burgos2007nestedness}, but the origin of nestedness in these networks remains an open question~\cite{payrato2019breaking}. Furthermore, the nested architecture of networks has been shown to be positively correlated with structural robustness (studied for instance through the lens of species extinction in ecology) when it is assumed that species with lower degree are more at risk of extinction, meaning that nested networks are maximally robust when the least linked species (specialists) become extinct but more fragile when the most linked species (generalists) face systematic extinction~\cite{burgos2007nestedness}. Following up on these findings,~\cite{rohr2014structural} have shown that maximally nested networks maximize the structural stability of mutualistic systems and that most observed networks were close to this optimum architecture.

Nestedness measures have been widely studied as one of the key metrics characterizing interactions in bipartite networks. Measuring it, however, has been a topic of ongoing investigations~\cite{payrato2020measuring, bruno2020ambiguity, almeida2008consistent, ulrich2009consumer}, and a number of different methods have been developed~\cite{mariani2019nestedness}. Among the latter, several nestedness metrics such as the Atmar \& Patterson temperature or the overlap and decreasing fill (NODF)~\cite{payrato2020measuring} do not take into account the quantitative nature of the interaction matrix, reducing it to a binary interaction matrix. Since the nature of our data allows us to access detailed information about the frequency of interaction between nodes, link weights span several orders of magnitudes and we therefore opt for the spectral radius $\rho$~\cite{staniczenko2013ghost} metric, a nestedness measure that can precisely handle weighted networks. This is of particular relevance as it has been shown that networks could be thought to be nested in binary form, but were eventually found not to be nested when accounting for weighted interactions~\cite{staniczenko2013ghost}.

\subsubsection{Bipartite modularity}
\label{subsec:modularity}
The modularity $Q$ of a network is a structural measure of how frequently nodes in defined subgroups of the network (modules) interact with each other compared to their frequency of interactions with nodes of other subgroups. The adjacency matrix of a modular network thus presents blocks of dense interactions between nodes of a given subgroup, and few links with nodes of other subgroups. Here, we use a modularity measure developed specifically to take into account the bipartite nature of the network of study~\cite{beckett2016improved}.

\subsubsection{Connectance}
\label{subsec:connectance}
The connectance of a network is defined as the number of realized links divided by the number of possible links in the network. This structural metric has been shown to be linked to network complexity, degree distribution and network stability~\cite{pimm1979structure,may1972will}. Furthermore, in bipartite ecological networks, the level of connectance of the network has been shown to impact the relationship between modularity and nestedness~\cite{fortuna2010nestedness}, with low connectance networks displaying a positive correlation between modularity and nestedness and networks with a high connectance value displaying a negative correlation between modularity and nestedness.

\section{Results}

\subsection{Investor communities}
Starting from $2017$ investors, we apply the clustering methodology described in the methods section. We obtain 16 investor communities (Fig.~\ref{fig:networks}A) described in Table~\ref{tab:namesInvestors} in the appendix. The communities are relatively few in number and are heterogeneous in size (the smallest one, \textit{C.13}, is comprised of 19 investors and the largest, \textit{C.04}, is comprised of 239 investors). Community \textit{C.00} is composed of investors specialized in the \textit{Health Care} sector, \textit{C.01} of historic investors that have been active relatively homogeneously throughout the whole period of study , \textit{C.02} of generalist investors capable of investing different amounts at different stages without displaying a strong sectoral specialization, \textit{C.03} of early-stage cryptocurrency investors that started being active around 2020, \textit{C.04} of United Kingdom (UK) and Germany (DE)-focused investors, \textit{C.05} of late-stage and private equity (PE) investors, \textit{C.06} of early-stage and business angels (BAs), \textit{C.07} of a specific type of very early-stage investors called \textit{accelerators}, \textit{C.08} of Canada-focused early-stage investors and incubators, \textit{C.09} of France-focused investors with a slight preference for the \textit{Health Care} sector, \textit{C.10} of China-focused investors, \textit{C.11} of early-stage actors that started being active around 2013, \textit{C.12} of investors that started being active in 2014 capable of investing throughout all stages of the VC cycle, \textit{C.13} of Latin America (Brazil, Mexico, Colombia)-focused investors, \textit{C.14} of India and Southeast Asia (SEA)-focused investors (India, Indonesia, Singapore), \textit{C.15} of Japan-focused investors.

These investor communities present relatively straightforward identities, with some of them being mainly defined by their geography of investments (e.g. \textit{C.10} and \textit{C.14}), others being defined by their sectors of investment (e.g. \textit{C.00} and \textit{C.03}), others by their stage of investment (e.g. \textit{C.05} and \textit{C.07}), others by their temporal patterns of investment (e.g. \textit{C.01} and \textit{C.11}) and others by a combination of several dimensions (e.g. \textit{C.08} with a mix of the stage and geographical dimensions and \textit{C.09} with a mix of sectoral and geographical dimensions). Our previous work~\cite{carniel2023novel} has shown these clustering results to be robust to the decimation of the characteristic dimensions used to compute the similarities.

\subsection{Patent clusters}
\begin{figure*}[ht]
    \centering
    \includegraphics[width=18cm]{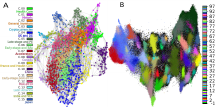}
    \caption{\textbf{A.} Pruned investor similarity network. Each node corresponds to an investor, and its color corresponds to the investor community it is allocated to. All investors can be grouped into 16 communities that define types of "investor species" (C.00 to C.15, presented in Fig.~\ref{fig:bipartite_graph} and Table~\ref{tab:namesInvestors}). \textbf{B.} Projected latent space (2 dimensions) of the patent data. Each point represents a patent and its color corresponds to its cluster allocation. The clustering defines 99 clusters, 98 thematic clusters and one unlabeled cluster. Cluster -1 (in black) corresponds to unlabeled data points.}
    \label{fig:networks}
\end{figure*}

Individual patents are grouped into clusters (Fig.~\ref{fig:networks}B), and we extract keywords and apply labels to describe the resulting patent clusters in Table~\ref{tab:namespatents} in the appendix. The size of the patent clusters is strongly heterogeneous, with the smallest patent cluster containing 226 patents and the largest cluster containing $82\ 513$ patents. Our patent clusters cover a wide range of specific technologies, the top 5 clusters by number of patents being : cluster $53$ (\textit{Pharmaceutical compositions/therapy}) with $82\ 513$ patents, cluster $41$ (\textit{Wireless Communication Technology}) with $51\ 139$ patents, cluster $55$ (\textit{Image Processing \& Autonomous Vehicles}) with $41\ 507$ patents, cluster $4$ (\textit{Pharmaceutical Compound Therapy}) with $30\ 981$ patents and cluster $64$ (\textit{Semiconductor Device Fabrication}) with $20\ 700$ patents. Examples of other patent clusters include cluster $12$ (\textit{Seismic Survey Techniques}, $322$ patents), cluster $38$ (\textit{Nucleic Acid Analysis}, $14\ 600$ patents) and cluster $15$ (\textit{Social Media Content}, $424$ patents). 

Technologies can be thought of as roughly being linked to three overarching groups : hardware-based, Health Care-related and software-based, each with its own specific challenges and constraints. Patented technologies can of course draw elements from several of these general fields, but we manually allocated each patent cluster to one of the three groups based on their label and keywords. We then colored the patent cluster nodes in Fig.~\ref{fig:bipartite_graph} following this allocation : \textit{Manufacturing} in green ($38$ patent clusters), \textit{Information Technology} (IT) in blue ($35$ patent clusters), and \textit{Health Care} in red ($24$ patent clusters). Even though there is some heterogeneity in the number of patent clusters in each group, all three groups are well-represented.

\subsection{Investor-patent network}
\begin{figure*}[ht]
    \centering
    \includegraphics{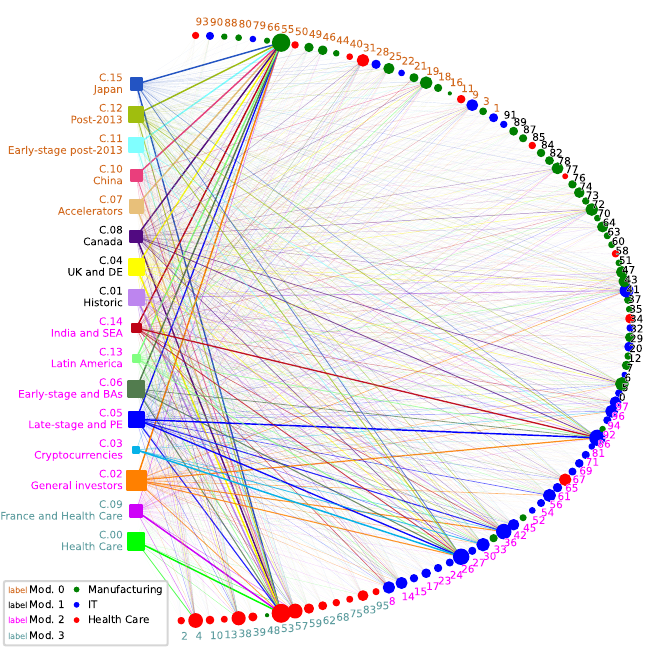}
    \caption{The investor community-patent cluster bipartite network. Square nodes represent investor communities and circle nodes patent clusters. Node sizes are a function of the node degrees. Link weights are normalized for each investor community by the maximum edge weight of the investor community, and the edge width shown is the logarithm of the normalized weight. A brief description of investor communities is provided under each investor community label, and a more extensive description is available in Table~\ref{tab:namesInvestors} in the appendix. Nodes were positioned following the 4 modules obtained by the bipartite modularity algorithm, and node label colors correspond to the module they were allocated to. Patent clusters are colored following a manual allocation of the high-level technological field they deal with (red for \textit{Health Care}, blue for \textit{Information Technology}, green for \textit{Manufacturing}).}
    \label{fig:bipartite_graph}
\end{figure*}

A bipartite network linking investor communities and patent clusters (Fig.~\ref{fig:bipartite_graph}) is built using investor communities and patent clusters. The degree distribution of the network is highly heterogeneous (truncated power law) both for investor communities and patent clusters, meaning that a small number of nodes have a large number of connections to other nodes while most others have a low number of connections. Broad-scale networks (networks exhibiting a truncated power law) are commonly found in abiotic and biotic systems, and are the result of finite size effects of the studied underlying network. The tail of the distribution (nodes with high degree) by definition has few observations, and as real processes are often bounded by the constraints of the system (in our case, a finite number of funding events), a bounded distribution is better suited to the system. 

The biadjacency matrix associated with the bipartite network is reordered following the descending node degree on both investor community and patent cluster nodes (upper-left packing). Community \textit{C.02} is the most active investor community, with a fairly diversified patent portfolio. Community \textit{C.00} is the second most active community, with a strong specialization in \textit{Health Care}-related patent clusters (clusters 53, 57, 4 and 38). On the other end of the matrix, communities \textit{C.13} and \textit{C.03} are the least active, with \textit{C.13} showing no specific pattern and \textit{C.03} showing IT and finance-related patent activity (clusters $26$, $36$, $30$ and $96$). The nested interaction pattern of the network is visible, with a strong density of interaction in the upper-left corner of the matrix and few interactions in the lower-right corner of the matrix. We visually observe that the nested structure, when rearranging the biadjacency matrix by descending order of degree, is imperfect in part due to the specificity of community \textit{C.00}. Indeed, since we work with quantitative rather than binary data, it boasts both a high degree (high number of interactions) and a high specialization (relatively few patent clusters with which it interacts).

\subsection{Connectance}
The measured connectance of our network is $C = 0.72$, a high value compared to ecological bipartite networks. As there are in theory no forbidden interactions (interactions that are structurally impossible in a network for physiological or phenological reasons) in our system and as our study covers a long period of time, this high value is not surprising. The magnitude of the connectance has strong implications on other structural network metrics such as degree distribution, nestedness and modularity.

\subsection{Modularity}
Using a modularity-based community detection algorithm, we measure the modularity value of our network and retrieve $4$ modules. The measured modularity of our network is $Q_{m} = 0.19$, meaning that the network is weakly modular (modularity ranges from $-1$ to $1$, with negative values corresponding to anti-modular networks and positive values to modular networks). The $4$ retrieved modules are shown by text color for patent cluster nodes and investor communities in Fig.~\ref{fig:bipartite_graph}. Three of these modules show a strong technological focus (module $1$ around \textit{Manufacturing}, module $2$ around \textit{Information Technologies} and module $3$ around \textit{Health Care}), with the fourth one (module $0$) containing a mix of technologies. We also computed the normalized modularity $\overline{Q}$ of a number of bipartite ecological networks and compared it to the normalized bipartite modularity of our network $\overline{Q}_{m}$ (results not shown here). We find that our modularity is lower than most networks it was compared to, potentially due to the different underlying nature of this socio-economic network compared to ecological networks.

\subsection{Relevance tests and ecological metrics}
\begin{figure*}[ht]
    \centering
    \includegraphics[width=\textwidth]{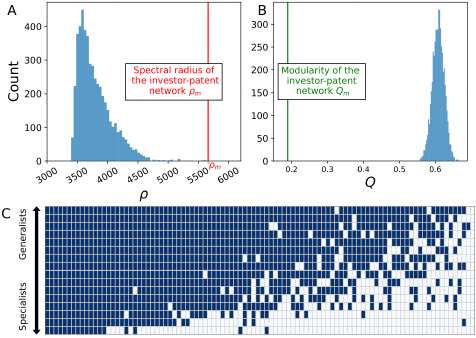}
    \caption{Statistical relevance tests for the nestedness and the modularity of the network.
    \textbf{(A)} Statistical relevance test for the nestedness $\rho_{m}$ (red vertical line) of the investor community-patent cluster network compared with $5\ 000$ iterations of a null model (blue histogram) keeping the qualitative interaction structure but shuffling the quantitative interaction structure. We see that our network is significantly more nested compared to networks generated by the null model. 
    \textbf{(B)} Statistical relevance test for the modularity $Q_{m}$ (green vertical line) of the investor community-patent cluster network compared with $5\ 000$ iterations of the null model (blue histogram). We see that our network is significantly less modular compared to networks generated by the null model.
    \textbf{(C)} Binarized representation of the biadjacency matrix. Investor communities correspond to the rows, patent clusters to the columns. The rows and columns are reordered by descending marginals (sums of the value of the row or column), yielding an upper-left packed matrix. The nested structure is displayed, with more generalist investor communities (bottom rows of the matrix) interacting only with a subset of the patent clusters the generalist species (top rows of the matrix) interact with.}
    \label{fig:metrics}
\end{figure*}

Relevance tests for the nestedness and modularity of our network are performed, and the investor community-patent cluster network is found to be significantly more nested (Fig.~\ref{fig:metrics}A, $\rho_{m} = 5662$, $\rho_{null} = 3799 \pm 295$, mean $\pm$ std, $z_{\rho} = 6.32$) and significantly less modular (Fig.~\ref{fig:metrics}B, $Q_{m} = 0.19$, $Q_{null} = 0.61 \pm 0.02$, $z_{Q} = -21$) compared to the null model. This specific network topology has strong implications on the properties of the network, notably in terms of robustness to external perturbations such as species extinction and in terms of species diversity.
The statistically high nestedness and low modularity (compared to the null models) of the interaction structure between investor communities and patent clusters is in line with previous findings in the literature as nestedness and modularity have been shown to be anticorrelated for networks with high connectance~\cite{fortuna2010nestedness}.
We also perform this analysis on a network where interactions are weighted by total funding amounts rather than number of interactions, and, even though the ordering of investor communities by degree is different, we find similar results in terms of nestedness and modularity (results not shown here).

\section{Discussion}

We observe that the bipartite network of interactions between startup investors and the patents in which they indirectly invest exhibits an emergent topological mutualism. This mutualistic topology, commonly found in ecological systems, results here from the presence of many investors whose generalist nature is induced by their portfolio diversification strategies, on one side of the network, and of general-purpose technologies with a broad spectrum of use, on the other. On the investor side, portfolio diversification is a fundamental and basic idea of modern portfolio theory~\cite{markowitz1952}. Investors are statistically better off if and when they diversify the risks they take among their investments: typically here, by supporting startups that develop different kinds of technological innovations. On the other side of the network, general purpose technologies~\cite{bresnahan1995general} are technologies characterized by their pervasiveness because they are used as inputs by many downstream sectors: here, by numerous startups with many different products that benefit from investments from many different types of investors. Although we do not address here the complexity of economies properly speaking and notably trade networks~\cite{bustos2012dynamics}, this result is closely related to the studies that have shed new light on the workings of socio-economic systems within the framework of economic complexity, and shares with them the rationale according to which the understanding of many societal issues implies to look at the systemic interactions that produce them~\cite{hidalgo2021economic,balland2022new}. In addition, but with a less pronounced relevance with respect to our approach, it should also be noted that another line of investigation has also attempted to draw parallels between the study of mutualistic systems and economic issues by introducing economic market effects to explain the evolution and stability of mutualistic interactions in ecological systems~\cite{noe1995biological, bshary2023marine}.

Topologically mutualistic networks have been shown to be significantly nested~\cite{bascompte2003nested}, a property that has been related to network robustness both for socio-economic systems and in ecology. Both literatures concur that the observed nested structure of the bipartite matrices describing topologically mutualistic networks contribute to their robustness and stability (e.g. ~\cite{hernandez2018trust,mariani2019nestedness}) and study the vulnerabilities of such systems, with the general conclusion that a nested system reacts very differently to perturbations depending on which types of nodes they affect.

Studies of the bipartite network of interactions between designer and contractor firms in the New York City garment industry~\cite{saavedra2009simple,saavedra2011strong}, following Uzzi's seminal work~\cite{uzzi1996sources} on the sources and consequences of embeddedness~\cite{granovetter1985problem}, have typically highlighted the fact that since the nested architecture of mutualistic networks implies that nodes contribute heterogeneously to their vulnerability, the removal of some nodes that specially contribute to the global nestedness of the network is consequentially associated with stronger vulnerabilities. In a similar vein, ~\cite{hernandez2018trust} have studied the Boulogne-sur-Mer Fish Market, focusing on the bipartite interactions between buyers and sellers, and studied its resistance to perturbations that would affect high-degree buyers or sellers, with the conclusion that the auction part of this market was more robust. The theory of economic complexity has concurrently associated nestedness and the dynamics of industrial ecosystems~\cite{bustos2012dynamics} notably in relation to the resilience of economies to external shocks~\cite{hausmann2011network,balland2022new}.

In ecology, where the nested network structure tends to minimize competition between species and support greater biodiversity~\cite{bastolla2009architecture}, maximizing structural stability~\cite{rohr2014structural}, perturbations impacting generalist species have been shown to lead to faster species depletion at the network level~\cite{burgos2007nestedness} by isolating specialist species due to the nested structure.

Furthermore, and with respect to the stability of nested bipartite networks, ecological studies have further shown that nested interaction networks emerge by considering an optimization principle aimed at maximizing species abundance~\cite{suweis2013emergence} and that nested mutualistic interactions boost equilibrium population densities and increase the resilience of communities~\cite{stone2020stability} : typically, when analyzing the short-term dynamics following a strong population perturbation, mutualist networks are associated with an ability to replete affected communities when species numbers fall dangerously low~\cite{stone2020stability}.

With respect to the investor-patent network, the attrition of generalist communities of investors should therefore be associated with a potentially severe impact on the entire network, an impact that could put technological fields and the associated emerging technologies at risk. Perturbations targeting generalist investors such as communities \textit{C.02} (investors capable of investing different amounts at different stages without a strong sectoral specialization) and \textit{C.06} (early-stage and business angels) stand out as the highest vulnerabilities, which highlights the role that these investors play for the system as a whole and for the diversity of technological fields of innovation that receive funding. In addition to these two communities, using funding amounts to weight the bipartite network links instead of the number of interactions, community \textit{C.05} (late-stage and private equity investors) also stands out as the most generalist investor community in this nested bipartite network, an observation which could be of special relevance since this community has been subject to a decrease in activity since 2022~\cite{pitchbook2023vc, cnbc2023softbank, ft2022mauling}. Such a perturbation, as it affects a community crucial to the nested patent-investor network, could be expected to have not only quantitative consequences, as is generally foreseen, but also qualitative ones, putting fields of technological innovation at risk.

In parallel, investor communities such as \textit{C.00} (investors specialized in the \textit{Health Care} sector) and \textit{C.03} (early-stage cryptocurrency investors) act as specialist investors in the bipartite network, whose emergence could be related to the need for specific skillsets in these sectors: for instance, the Health Care sector (that corresponds here to module 3 in Fig.~\ref{fig:bipartite_graph}) is well-known to be associated with very specific regulation and R\&D cycles.

A relevant question here is whether the number of technological and sectoral specificities will increase in relation to the development of so-called deep techs, a category that includes medtechs, quantum computing or artificial intelligence, each of which is associated with specific and emerging regulations. Such a phenomenon could lead to an increased number of specialist investor communities, which would in turn increase the modularity of the patent-investor network, a phenomenon negatively correlated with network robustness as mentioned above but which could also mitigate perturbations affecting key generalist investors. A more modular structure of the bipartite patent-investor network could result in fields of technological innovations being dependent on a limited number of investor communities for their funding, but conversely less dependent upon generalist investors.

Finally, and with respect to future research, startup databases are known to under-represent early-stage funding rounds compared to later-stage ones due to an easier tracking of the latter. Although we do not expect such a bias to affect the results of our study, complementary analyses on reduced but more exhaustive datasets could further clarify this issue. In addition, VC funding tends to target novel, emerging and potentially disruptive technologies, while others are funded by a more varied panel of investments which could also warrant more comprehensive investigations, notably innovations that spin-off from academia in the context of its specific set of institutions and incentives~\cite{dasgupta1994toward}.
Further studies could also attempt to directly assess the robustness of the investor-startup network with respect to technological diversity~\cite{hidalgo2009building, zhang2023cultivated} when facing different types of crises.

\section{Conclusion}
In this work, by leveraging large-scale financial, startup and patent datasets, we have built a bipartite network directly linking investors and technologies. Using network metrics, we have found this network to display an emergent topological mutualism, associated with a heterogeneous degree distribution, a significant nestedness and a significantly lower modularity compared to null models. This has relevant implications for the robustness of the ability of startups and investors to support technological innovation when facing crises. We notably expect the system to react differently depending on perturbations. In particular, perturbations affecting investor communities that contribute strongly to the nestedness of the patent-investor network could have a far-reaching impact on technological innovation.

\bibliographystyle{unsrt}  
\bibliography{references}  

\clearpage

\vspace{2cm}

{\bf \Huge Appendix}

\vspace{2cm}

\begin{table*}[h]
    \centering%
    \resizebox{0.99\textwidth}{!}{%
    \begin{tabular}{c|c|c}
        Investor community & \# of investors & Brief community description \\
         \hline
        C.00 & 110 & Health Care investors \\
         \hline
        C.01 & 132 & Historic investors \\
         \hline
        C.02 & 200 & Generalist investors active whole period \\
         \hline
        C.03 & 27 & Cryptocurrency investors \\
         \hline
        C.04 & 239 & EU-focused investors (UK and DE) \\
         \hline
        C.05 & 102 & Late-stage investors and PE \\
         \hline
        C.06 & 236 & Early-stage and BAs \\
         \hline
        C.07 & 189 & Accelerators \\
         \hline
        C.08 & 80 & Canada-focused investors \\
         \hline
        C.09 & 71 & France-focused Health Care-focused investors \\
         \hline
        C.10 & 122 & China-focused investors \\
         \hline
        C.11 & 158 & Early-stage post-2013 investors \\
         \hline
        C.12 & 201 & "New-generation" post-2013 investors\\
         \hline
        C.13 & 19 & Latin America-focused investors\\
         \hline
        C.14 & 78 & India and SEA-focused investors\\
         \hline
        C.15 & 53 & Japan-focused investors\\
    \end{tabular}
    }
    \caption{Description of investor communities. \textbf{UK} stands for \textit{United Kingdom}, \textbf{DE} for \textit{Germany}, \textbf{PE} for \textit{Private Equity}, \textbf{BA} for \textit{Business Angel}, \textbf{SEA} for \textit{Southeast Asia}.
    "Historic" investors are investors that have been active for a long period of time, since the late 1990s-early 2000s.
    "Generalist" investors are investors that do not display a significant sectoral focus, investing in all types of sectors and related technologies.
    "Cryptocurrency" investors are investors strongly specialized in cryptocurrencies and related financial sectors.
    "Late-stage" investors focus on the later stages of VC financing (series B and onwards), typically investing very large amounts.
    "Early-stage" investors focus on early stages of VC financing (pre-seed, seed and series A), investing relatively small amounts.
    "Business Angels" are individuals who invest their own money in startups, usually in early-stage rounds and low amounts.
    "Accelerators" are a specific type of early-stage investors that usually operate by selecting batches of companies for a short period, providing them with small amounts of money and an intensive mentoring program of a few months focused on developing specific aspects of the company.
    "Post-2013" investors are investors that started being active (or greatly increased their activity) around the 2013 period, where VC financing experienced sudden and significant growth.}
    \label{tab:namesInvestors}
\end{table*}

\begin{table}[]
\centering%
\SetTblrInner{rowsep=0pt}
\resizebox*{0.99\textwidth}{!}{%
\begin{tblr}{c|c|c|c|>{\centering}m{6cm}}
\textbf{patent cluster} &
  \textbf{\# of connections} &
  \textbf{cluster label} &
  \textbf{1-grams} &
  \textbf{2-grams} \\ \hline
0 &
  74 &
  Video Displaying Technology &
  video | format | stream | frames | media &
  video stream | video data | video content \\ \hline
1 &
  186 &
  Location-based Wireless Technology &
  location | wireless | positioning | mobile | satellite &
  mobile device | mobile station | wireless device \\ \hline
2 &
  62 &
  Cancer Treatment Therapies &
  cancer | treating | inhibitor | combination | treatment &
  treating cancer | methods treating | combination therapy \\ \hline
3 &
  77 &
  Fluid Valve Assembly &
  valve | piston | fluid | chamber | pressure &
  valve assembly | pressure tube | shock absorber \\ \hline
4 &
  2801 &
  Pharmaceutical Compound Therapy &
  compounds | formula | thereof | derivatives | diseases &
  pharmaceutical compositions | pharmaceutically acceptable | compounds formula \\ \hline
5 &
  1938 &
  Power Electronics Circuit &
  power | voltage | circuit | output | signal &
  power supply | clock signal | input signal \\ \hline
6 &
  15 &
  Wireless Network Technology &
  network | wireless | access | mobile | service &
  wireless network | network access | wireless device \\ \hline
7 &
  232 &
  Chemical Reaction Engineering &
  catalyst | process | gas | stream | carbon &
  gas stream | carbon dioxide | stream comprising \\ \hline
8 &
  1104 &
  Multimedia Streaming Services &
  media | content | video | playback | audio &
  media content | video content | playback device \\ \hline
9 &
  862 &
  Speech Processing Technology &
  audio | speech | sound | voice | microphone &
  audio signal | audio data | speech recognition \\ \hline
10 &
  149 &
  Pharmaceutical Formulations \& Dosage &
  pharmaceutical | formulations | composition | release | oral &
  pharmaceutical composition | pharmaceutically acceptable | dosage form \\ \hline
11 &
  132 &
  Microbial Acid Production &
  acid | production | microorganisms | microbial | amino &
  amino acid | method producing | non naturally \\ \hline
12 &
  60 &
  Seismic Survey Techniques &
  seismic | sensor | measurement | subsurface | acoustic &
  seismic data | seismic trace | acoustic signal \\ \hline
13 &
  41 &
  Virus, Vaccine, Antigen &
  virus | vaccine | protein | recombinant | vectors &
  present invention | invention relates | nucleic acid \\ \hline
14 &
  966 &
  Online Advertising Services &
  advertisement | advertising | ad | campaign | advertiser &
  advertising campaign | advertising content | web page \\ \hline
15 &
  119 &
  Social Media Content &
  content | item | social | user | online &
  content item | social networking | social media \\ \hline
16 &
  1 &
  Vehicle Braking Systems &
  brake | disc | caliper | disk | lining &
  disc brake | brake disc | brake caliper \\ \hline
17 &
  266 &
  Cloud Computing Services &
  application | computing | cloud | software | service &
  cloud computing | computing environment | virtual machine \\ \hline
18 &
  106 &
  Augmented Reality Displays &
  light | display | pixel | image | eye &
  display device | light emitting | image light \\ \hline
19 &
  1247 &
  Motor Vehicle Assembly &
  rotor | motor | shaft | assembly | vehicle &
  steering column | aerial vehicle | motor vehicle \\ \hline
20 &
  363 &
  Video Compression Technology &
  video | block | coding | prediction | picture &
  video data | video coding | motion vector \\ \hline
21 &
  200 &
  Agricultural Management \& Yield &
  crop | agricultural | yield | plant | field &
  agricultural field | crop yield | management zones \\ \hline
22 &
  31 &
  Location-based Tracking Technology &
  location | mobile | tracking | geo | fence &
  mobile device | tracking device | location information \\ \hline
23 &
  111 &
  Multi-Tenant Database &
  database | application | custom | tenant | object &
  access permissions | multi tenant | mechanisms methods \\ \hline
24 &
  91 &
  Network Management \& Security &
  network | traffic | service | policy | proxy &
  communication network | network element | network agent \\ \hline
25 &
  625 &
  Battery Electrochemistry Technology &
  fuel | electrolyte | battery | anode | electrode &
  fuel cell | lithium ion | active material \\ \hline
26 &
  4869 &
  Payment \& Transaction systems &
  payment | transaction | merchant | card | account &
  point sale | systems methods | mobile device \\ \hline
27 &
  102 &
  Web Page Management &
  web | page | content | tab | user &
  web page | context menu | tabs tab \\ \hline
28 &
  220 &
  Neural Network Technology &
  neural | training | network | input | output &
  neural network | convolutional neural | training neural \\ \hline
29 &
  314 &
  Ultrasound Medical Imaging &
  ultrasound | imaging | tissue | image | ultrasonic &
  ultrasound imaging | ultrasound device | ultrasound data \\ \hline
30 &
  1713 &
  Identity Authentication Technology &
  authentication | key | identity | user | server &
  user authentication | public key | private key \\ \hline
31 &
  1204 &
  Medical Monitoring Devices &
  physiological | patient | monitoring | heart | blood &
  heart rate | blood pressure | vital signs \\ \hline
32 &
  68 &
  Software Test Platform &
  application | software | platform | test | file &
  live multi | multi tenant | sdk platform \\ \hline
33 &
  111 &
  Footwear Assembly Tools &
  tubular | upper | footwear | portion | projectile &
  tubular element | projectile casing | entangling projectile \\ \hline
34 &
  285 &
  Polymer Composition Formulations &
  composition | polymer | weight | comprising | containing &
  composition comprising | invention relates | present invention \\ \hline
35 &
  32 &
  Heat Dissipation Technology &
  heat | cooling | air | thermal | coolant &
  heat dissipation | heat sink | heat exchanger \\ \hline
36 &
  3542 &
  Data Storage Systems &
  storage | memory | cache | file | data &
  data storage | encoded data | dispersed storage \\ \hline
37 &
  102 &
  HVAC Climate Control &
  temperature | hvac | thermostat | energy | setpoint &
  energy consumption | setpoint temperature | ambient temperature \\ \hline
38 &
  2298 &
  Nucleic Acid Analysis &
  nucleic | sample | acid | dna | sequencing &
  nucleic acid | methods compositions | invention provides \\ \hline
39 &
  189 &
  Medical Neural Stimulation &
  stimulation | nerve | tissue | electrical | electrode &
  electrical stimulation | nerve stimulation | peripheral nerve \\ \hline
40 &
  30 &
  Patient Support Equipment &
  support | patient | deck | frame | foot &
  patient support | support apparatus | hospital bed \\ \hline
41 &
  2063 &
  Wireless Communication Technology &
  wireless | ue | station | transmission | channel &
  base station | wireless communication | user equipment \\ \hline
42 &
  723 &
  Social Networking Platform &
  social | networking | users | online | content &
  social networking | user social | online social \\ \hline
43 &
  977 &
  Steel Cutting/Coating &
  steel | material | sheet | surface | coating &
  steel sheet | method producing | cutting edge \\ \hline
44 &
  25 &
  Light Imaging Technology &
  light | image | imaging | lidar | sensor &
  image sensor | light field | light pulses \\ \hline
45 &
  43 &
  Electronic Connectors &
  connector | electrical | electronic | plug | housing &
  electrical connector | electronic device | connector includes \\ \hline
46 &
  280 &
  Magnetic Sensor Devices &
  sensor | magnetic | field | sensing | current &
  magnetic field | magnetic sensor | field sensor \\ \hline
47 &
  890 &
  Organic LED/Solar &
  solar | layer | light | emitting | photovoltaic &
  light emitting | solar cell | emitting device \\ \hline
48 &
  0 &
  Integrated Circuit Devices &
  circuit | transistor | voltage | integrated | semiconductor &
  integrated circuit | semiconductor integrated | circuit includes \\ \hline
49 &
  228 &
  3D Printing Technology &
  printing | ink | dimensional | build | printer &
  dimensional printing | imprint lithonetworky | print head \\ \hline
\end{tblr}
}
\end{table}

\begin{table}[]
\centering%
\SetTblrInner{rowsep=0pt}
\resizebox*{0.99\textwidth}{!}{%
\begin{tblr}{c|c|c|c|>{\centering}m{6cm}}
\textbf{patent cluster} &
  \textbf{\# of connections} &
  \textbf{cluster label} &
  \textbf{1-grams} &
  \textbf{2-grams} \\ \hline
50 &
  62 &
  Food Composition \& Protein &
  protein | food | composition | soy | product &
  soy protein | protein solution | food product \\ \hline
51 &
  37 &
  Solar Energy Conversion &
  solar | photovoltaic | dc | power | inverter &
  photovoltaic power | dc power | solar power \\ \hline
52 &
  32 &
  Content Delivery Network &
  cdn | content | delivery | server | origin &
  content delivery | delivery network | network cdn \\ \hline
53 &
  10004 &
  Pharmaceutical compositions/therapy &
  compositions | invention | acid | cells | present &
  present invention | invention relates | invention provides \\ \hline
54 &
  95 &
  Web Content Management &
  content | folder | web | collection | item &
  collection folder | content management | content item \\ \hline
55 &
  8393 &
  Image Processing \& Autonomous Vehicles &
  image | vehicle | object | display | camera &
  image data | aerial vehicle | unmanned aerial \\ \hline
56 &
  1464 &
  Cybersecurity \& Threat Detection &
  security | malware | threat | malicious | risk &
  malware detection | security platform | anomalies threats \\ \hline
57 &
  2954 &
  Medical Devices and Implants &
  distal | catheter | implant | tissue | end &
  distal end | proximal end | devices methods \\ \hline
58 &
  67 &
  Medical Stimulation Devices &
  stimulation | tissue | ultrasound | nerve | transcutaneous &
  transcutaneous stimulation | adipose tissue | electrical stimulation \\ \hline
59 &
  386 &
  Pharmaceutical Treatment Methods &
  treating | treatment | administering | disease | compositions &
  methods treating | compositions methods | pharmaceutically acceptable \\ \hline
60 &
  41 &
  Fluid Management Systems &
  flow | valve | tubular | pipe | fluid &
  tubular element | tubular section | flow path \\ \hline
61 &
  234 &
  Electronic Messaging Platform &
  message | notification | messaging | user | email &
  electronic message | agent performance | contact information \\ \hline
62 &
  175 &
  Drug Delivery Devices &
  needle | drug | dose | syringe | delivery &
  delivery device | drug delivery | piston rod \\ \hline
63 &
  47 &
  Optical Imaging and Analysis &
  radiation | imaging | ray | detector | optical &
  absorption data | light source | radiation source \\ \hline
64 &
  576 &
  Semiconductor Device Fabrication &
  semiconductor | layer | substrate | region | gate &
  semiconductor device | semiconductor substrate | dielectric layer \\ \hline
65 &
  1104 &
  Healthcare Information Systems &
  patient | health | medical | healthcare | care &
  health care | medical information | patient data \\ \hline
66 &
  18 &
  Acoustic-Piezoelectric Devices &
  piezoelectric | acoustic | resonator | idt | surface &
  acoustic wave | piezoelectric plate | piezoelectric element \\ \hline
67 &
  94 &
  Messaging \& Collaboration &
  message | messaging | user | chat | conversation &
  instant messaging | user device | electronic message \\ \hline
68 &
  62 &
  Stem Cell Research &
  cells | stem | pluripotent | progenitor | differentiation &
  stem cells | pluripotent stem | progenitor cells \\ \hline
69 &
  145 &
  Gambling Games &
  game | wager | player | gambling | entertainment &
  entertainment game | real world | hybrid game \\ \hline
70 &
  31 &
  Internal Combustion Engine &
  piston | valve | damper | crankshaft | engine &
  connecting rod | control valve | compression ratio \\ \hline
71 &
  108 &
  Media Content Recommendation &
  content | item | media | user | ratings &
  content item | media content | content based \\ \hline
72 &
  1275 &
  Fluid \& Gas Systems &
  gas | fluid | liquid | chamber | air &
  exhaust gas | heat exchanger | compressed air \\ \hline
73 &
  56 &
  Touch Sensing Technology &
  touch | sensing | capacitive | sensor | capacitance &
  touch sensor | touch sensitive | touch panel \\ \hline
74 &
  474 &
  Lidar Optical Technology &
  optical | light | lidar | laser | beam &
  light source | light beam | optical signal \\ \hline
75 &
  43 &
  Surgical Robotics Cluster &
  surgical | instrument | robot | tool | effector &
  surgical instrument | end effector | surgical tool \\ \hline
76 &
  129 &
  Biosensor Analysis Technology &
  sample | analyte | electrode | test | biosensor &
  working electrode | liquid sample | flow cell \\ \hline
77 &
  14 &
  Nucleic Acid Biotechnology &
  nucleic | rna | acid | expression | sequence &
  nucleic acid | control elements | promoter control \\ \hline
78 &
  878 &
  Optical Networking Technology &
  optical | light | laser | waveguide | wavelength &
  optical signal | optical fiber | light source \\ \hline
79 &
  27 &
  Gaming \& Accessory &
  game | player | battle | gaming | server &
  game content | server device | game program \\ \hline
80 &
  30 &
  Vehicle Tire Monitoring &
  tire | vehicle | sensor | pressure | door &
  tire pressure | pressure sensor | door lock \\ \hline
81 &
  28 &
  Television Program Guide &
  television | program | guide | interactive | schedule &
  program guide | television program | interactive television \\ \hline
82 &
  164 &
  Semiconductor Memory Devices &
  memory | semiconductor | layer | cell | bit &
  memory cell | memory device | semiconductor memory \\ \hline
83 &
  540 &
  Antibody Therapy Research &
  antibodies | antibody | binding | anti | cd &
  antigen binding | binding fragments | present invention \\ \hline
84 &
  127 &
  Measurement Sensor Technology &
  sensor | measuring | acoustic | measurement | gas &
  gas sensor | data sheet | tank floor \\ \hline
85 &
  61 &
  Fluid Pump Devices &
  fluid | pump | blood | gas | breast &
  breast pump | breathing gas | piezo air \\ \hline
86 &
  4816 &
  Search Engine Technology &
  search | query | document | results | queries &
  search results | search query | search engine \\ \hline
87 &
  107 &
  Fluid Analysis Technology &
  sample | fluid | flow | chamber | cartridge &
  flow cell | fluid sample | analytical instrument \\ \hline
88 &
  23 &
  Image Forming Devices &
  toner | sheet | member | forming | roller &
  image forming | forming apparatus | main body \\ \hline
89 &
  172 &
  Microfluidic Devices/Systems &
  microfluidic | fluid | sample | droplet | channel &
  microfluidic device | microfluidic channel | cell processing \\ \hline
90 &
  99 &
  Location Tracking Technology &
  location | mobile | devices | geonetworkic | determination &
  mobile device | location information | location data \\ \hline
91 &
  55 &
  Serial Bus Protocol &
  bus | serial | clock | signal | bit &
  serial bus | clock signal | digital data \\ \hline
92 &
  24 &
  Wireless Power Management &
  power | wireless | transmit | consumption | communication &
  power control | transmit power | power consumption \\ \hline
93 &
  49 &
  Fluid Dispensing Devices &
  dispensing | container | dispenser | outlet | liquid &
  dispensing apparatus | dispensing device | liquid medicine \\ \hline
94 &
  100 &
  User Interface Design &
  interface | user | ui | networkical | application &
  user interface | networkical user | computing device \\ \hline
95 &
  45 &
  RNAi Therapy Cluster &
  expression | compositions | gene | agents | treating &
  compositions methods | double stranded | invention provides \\ \hline
96 &
  1380 &
  Networking \& Traffic Management &
  network | packet | traffic | routing | node &
  network traffic | network device | virtual network \\ \hline
97 &
  850 &
  Telecommunication Services \& Devices &
  telephone | voice | caller | message | calls &
  telephone number | calling party | called party \\
\end{tblr}
}
    \caption{Names of the patent clusters according to their ngrams. Cluster labels were inferred from the top 20 1-grams and top 20 2-grams. The number of connections represents the number of connections between the patent cluster and all investor communities in the bipartite network.}
    \label{tab:namespatents}
\end{table}

\end{document}